\documentclass{article}
\usepackage{spconf,amsmath,graphicx}

\usepackage{import, color} %
\usepackage{siunitx}
\usepackage{multirow}
\usepackage{tabularx}
\usepackage{url}

\usepackage{caption}
\definecolor{LMSred}{rgb}{0.80,0.20,0.20}

\newcommand{\OME}{RME}
\newcommand{\RME}{RME}
\newcommand{\OMC}{RMC}
\newcommand{\FaOMC}{\mbox{FRMC}}
\newcommand{\FourNNC}{NNC}
\newcommand{\RFSR}{\mbox{R-FSR}}
\newcommand{\RFSRmodified}{\mbox{D-FSR}}

\ninept
\title{Novel Consistency Check for Fast Recursive Reconstruction of Non-Regularly Sampled Video Data}
\name{Simon Grosche, J\"urgen Seiler, and  Andr\'e Kaup }
\address{Multimedia Communications and Signal Processing\\ Friedrich-Alexander-Univerist\"at Erlangen-N\"urnberg, Cauerstr. 7, 91058 Erlangen, Germany\\ \textit{\small \{simon.grosche, juergen.seiler, andre.kaup\}@fau.de}}

\usepackage[firstpage=true]{background}
\usepackage{hyperref}

\SetBgContents{\parbox{\textwidth}{\footnotesize © 2021 IEEE. Personal use of this material is permitted. Permission from IEEE must be
		obtained for all other uses, in any current or future media, including
		reprinting/republishing this material for advertising or promotional purposes, creating new
		collective works, for resale or redistribution to servers or lists, or reuse of any copyrighted
		component of this work in other works. DOI: \href{https://doi.org/10.1109/ICIP42928.2021.9506567}{10.1109/ICIP42928.2021.9506567.}}}
\SetBgScale{1}
\SetBgAngle{0}
\SetBgPosition{current page.south}
\SetBgVshift{1cm}
\SetBgColor{black}
\SetBgOpacity{1}

\begin{document}

\maketitle

\begin{abstract}\vspace*{-0.5mm}
	Quarter sampling is a novel sensor design that allows for an acquisition of higher resolution images without increasing the number of pixels.
	When being used for video data, one out of four pixels is measured in each frame.
	Effectively, this leads to a non-regular spatio-temporal sub-sampling.
	Compared to purely spatial or temporal sub-sampling, this allows for an increased reconstruction quality, as aliasing artifacts can be reduced.
	For the fast reconstruction of such sensor data with a fixed mask, recursive variant of  frequency selective reconstruction (FSR) was proposed. Here, pixels measured in previous frames are projected into the current frame to support its reconstruction. In doing so, the motion between the frames is computed using template matching. Since some of the motion vectors may be erroneous, it is important to perform a proper consistency checking.
	In this paper, we propose  faster consistency checking methods as well as  a novel recursive FSR that uses the projected pixels different than in literature and can handle dynamic masks. Altogether, we are able to significantly increase the reconstruction quality by \SI[retain-explicit-plus]{+1.01}{dB}  compared to the state-of-the-art recursive reconstruction method using a fixed mask. Compared to a single frame reconstruction, an average gain of about \SI[retain-explicit-plus]{+1.52}{dB} is achieved for dynamic masks.  At the same time, the computational complexity of the consistency checks is reduced by a factor of 13 compared to the literature algorithm.
\end{abstract}
\begin{keywords}
Non-Regular Sampling, Image Reconstruction
\end{keywords}

\section{INTRODUCTION}
\begin{figure}[t]
	\centering
	{\footnotesize
		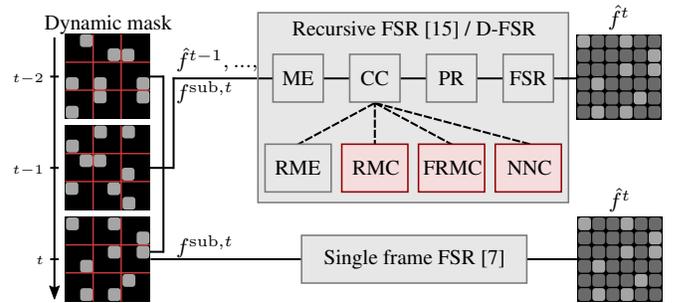
	}
	\vspace*{-4mm}
	\footnotesize

	\caption{Flow diagram of quarter sampling for video data using a dynamic mask. Our novel contributions to the consistency checks of recursive FSR are highlighted in red.  Abbreviations: ME: Motion estimation, CC: Consistency check, PR: Projection, \RME{}: Reverse motion estimation, \OMC{}: Reverse motion check, \FaOMC{}: Fast reverse motion check, \FourNNC{}: Nearest neighbor check.}
	\label{fig:flow_graph}
	\vspace*{-4mm}
\end{figure}

\vspace*{-2mm}
\label{sec:intro}
Using quarter sampling \cite{Schoberl2011}, the spatial resolution of an imaging sensor can be increased. This is achieved by physically covering three quarters of each pixel of a regular low-resolution sensor. Effectively, this leads to a non-regular sampling of the image with respect to a higher resolution grid with twice the resolution in both spatial dimensions as can be seen in Figure\,\ref{fig:flow_graph} (left).
Due to the non-regularity, visually disturbing aliasing artifacts that conventionally occur for regular sampling  can be reduced \cite{Dippe1985, Hennenfent2007, Maeda2009}.
For the reconstruction, frequency selective reconstruction (FSR) has shown to be a successful reconstruction scheme for various inpainting and extrapolation tasks \cite{Herraiz2008, Stehle2006} and gave best results for non-regular sampling and quarter sampling in~\cite{Schoberl2011,Seiler2015,Grosche2018}.
Quarter sampling, as well as any non-regular sub-sampling, can be seen as a special case of compressed sensing \cite{Candes2006, Donoho2006} as has been shown in \cite{Seiler2015, Grosche2020_localJSDE}. In the compressed sensing framework, the FSR can be interpreted as a special case of more general reconstruction algorithms from the class of matching pursuit algorithms \cite{Grosche2020_localJSDE,Mallat1993, Tropp2007}.

Besides still images, the acquisition of video data is of great importance. In combination with quarter sampling, video acquisition has been investigated for fixed quarter sampling masks \cite{Jonscher2015, Jonscher2016a} as well as for dynamic quarter sampling masks \cite{Jonscher2018}. For the latter, the sampling mask changes from frame to frame and a sophisticated read-out strategy is applied such that each pixel in a $2{\times}2$ block is read exactly once within four frames. Every fourth frame, the mask repeats.  Compared to a purely spatial or purely temporal sub-sampling, a more uniform placement of the pixels in time and space is achieved and a higher reconstruction quality is found \cite{Jonscher2018}. In this paper, we consider both fixed and dynamic masks.

Among those works, causal reconstruction algorithms such as the one in \cite{Jonscher2016a} are of special interest since they only use past measurements  for the reconstruction of the current frame. Such causal scenario is of special importance since future measurements are not available in real-time applications at the time of the reconstruction. To processing chain of the measurement an reconstruction is illustrated in Figure\,\ref{fig:flow_graph} for a current frame at time $t$ and two preceding frames.

In this paper, we focus on such causal reconstruction scenarios as it is done in \cite{Jonscher2016a}.
Our novel contributions to these scenarios are twofold:
As first contribution, we propose a novel combination of consistency checks that finds outliers among the motion vectors much faster and more reliable than in \cite{Jonscher2016a}. This is marked with red color in Figure\,\ref{fig:flow_graph}.
As second contribution, we propose the so called \RFSRmodified{} being a new implementation of recursive FSR that handles the projected pixel differently than in \cite{Jonscher2016a} and can be used with dynamic sampling masks. It is different  to \cite{Jonscher2016a}, where only fixed masks are considered and it is different from \cite{Jonscher2018}, where information from future frames is used.

Our analysis is performed on a variety of test sequences to show the wide applicability of the modifications. Besides analysis of the reconstruction quality of the proposed modifications, visual comparisons are provided, and the computation times are compared. 
This paper is organized as follows: In Section\,\ref{sec:state_of_art}, we present the state of the art. In Section\,\ref{sec:proposed_enhancements}, we describe our novel contributions. In Section\,\ref{sec:simulation_and_results}, the simulation results are presented and discussed. Section\,\ref{sec:conclusion} summarizes the paper.

\vspace*{-1mm}
\section{STATE OF THE ART}
\vspace*{-1mm}
\label{sec:state_of_art}
\vspace*{-1mm}
\subsection{Single Frame Reconstruction}
\vspace*{-1mm}
In order to reconstruct the missing pixels from the sampled image data, frequency selective reconstruction (FSR) has shown to provide high reconstruction quality outperforming other reconstruction techniques \cite{Seiler2015}.
The sub-sampled image can be understood as $f^\mathrm{sub}_{mn} = f_{mn} \cdot b_{mn}$, where $f$ and $b$ are the reference high resolution image and the binary mask. From the sub-sampled image and the mask,  FSR reconstructs an image $\hat{f}_{mn}$. It therefore subdivides the image into neighboring blocks that are  reconstructed using the measurements from their neighboring blocks, too. The model of each block is build in the Fourier domain where the image is assumed to be approximately sparse \cite{Lam2000, Elad2010}.

\vspace*{-1mm}
\subsection{Recursive FSR}
\begin{figure}[t]
	\centering
	{\footnotesize
\begingroup%
  \makeatletter%
  \providecommand\color[2][]{%
    \errmessage{(Inkscape) Color is used for the text in Inkscape, but the package 'color.sty' is not loaded}%
    \renewcommand\color[2][]{}%
  }%
  \providecommand\transparent[1]{%
    \errmessage{(Inkscape) Transparency is used (non-zero) for the text in Inkscape, but the package 'transparent.sty' is not loaded}%
    \renewcommand\transparent[1]{}%
  }%
  \providecommand\rotatebox[2]{#2}%
  \newcommand*\fsize{\dimexpr\f@size pt\relax}%
  \newcommand*\lineheight[1]{\fontsize{\fsize}{#1\fsize}\selectfont}%
  \ifx\svgwidth\undefined%
    \setlength{\unitlength}{244.69369507bp}%
    \ifx\svgscale\undefined%
      \relax%
    \else%
      \setlength{\unitlength}{\unitlength * \real{\svgscale}}%
    \fi%
  \else%
    \setlength{\unitlength}{\svgwidth}%
  \fi%
  \global\let\svgwidth\undefined%
  \global\let\svgscale\undefined%
  \makeatother%
  \begin{picture}(1,0.2819852)%
    \lineheight{1}%
    \setlength\tabcolsep{0pt}%
    \put(0,0){\includegraphics[width=\unitlength,page=1]{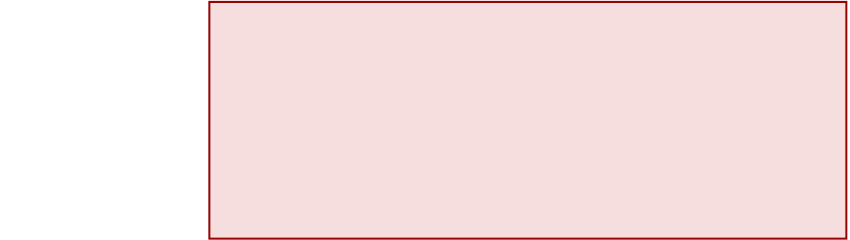}}%
    \put(0.37180808,0.24688168){\color[rgb]{0,0,0}\makebox(0,0)[t]{\lineheight{1.25}\smash{\begin{tabular}[t]{c}(b) \OMC{}\end{tabular}}}}%
    \put(0,0){\includegraphics[width=\unitlength,page=2]{explain_crosschecks.pdf}}%
    \put(0.6247601,0.24683255){\color[rgb]{0,0,0}\makebox(0,0)[t]{\lineheight{1.25}\smash{\begin{tabular}[t]{c}(c) \FaOMC{}\end{tabular}}}}%
    \put(0,0){\includegraphics[width=\unitlength,page=3]{explain_crosschecks.pdf}}%
    \put(0.8774292,0.24678369){\color[rgb]{0,0,0}\makebox(0,0)[t]{\lineheight{1.25}\smash{\begin{tabular}[t]{c}(d) \FourNNC{}\end{tabular}}}}%
    \put(0,0){\includegraphics[width=\unitlength,page=4]{explain_crosschecks.pdf}}%
    \put(0.11841988,0.2467835){\color[rgb]{0,0,0}\makebox(0,0)[t]{\lineheight{1.25}\smash{\begin{tabular}[t]{c}(a) Motion vectors\end{tabular}}}}%
  \end{picture}%
\endgroup%

	}
	\vspace*{-5mm}
	\footnotesize
	\caption{Illustration of (a) the motion vector field from frame $f^{(t)}$ to frame $f^{(t-1)}$  and  (b-d) the used consistency checks. In contrast to the values used in the text, \OMC{} is shown for a search range of $\{-2,-1,0,-1,2\}$ and \FaOMC{} is shown for a partial search range being $\{-2,0,2\}$.}
	\label{fig:consistency_checking}
	\vspace*{-4mm}
\end{figure}

\vspace*{-1mm}
For video data, it was proposed to additionally use information from previous frames \cite{Jonscher2016a}.  For any missing pixel in the current frame, a motion vector pointing to a measured pixel in one of the preceding frames could provide useful information.  In \cite{Jonscher2016a}, Jonscher et al. propose such an approach called recursive FSR (\RFSR{}). In \RFSR{}, the motion estimation is performed by a pixel-wise template matching using the already reconstructed past frames and the measured data from the current frame. Such motion vector field is illustrated in Figure\,\ref{fig:consistency_checking}\,(a).
During the template matching \cite{brunelli2009template}, some motion vectors may be untrustworthy. This can result from cases where the motion is larger than the search range, from occlusions or local optima. Since non-regularly sampled data is used, these issues are increased further as fewer information is available.

In order to sort out unfavorable motion vectors, Jonscher et al. propose a consistency check for which  the motion vectors in the reverse direction are calculated using an additional reverse motion estimation (\RME{}). Only when both motion vectors coincide, the motion vector is accepted. 
While this strategy is reasonable and seems successful, it is also computationally demanding since the number of cost functions that needs to be evaluated is doubled. 

With the accepted motion vectors at hand, values for some of the missing pixels the current frame can be found by following their motion vector into the past. If a measurement is available at the corresponding position in the past frame, it is projected to the current frame and it is used as an additional measurement during the reconstruction. In case projections from more than one past frame are available, these are averaged.
\section{Novel Contributions}
\vspace*{-1mm}
Our novel contributions to the recursive reconstruction of non-regularly sampled video data are twofold and described in the two following sub-sections.
\vspace*{-1mm}
\label{sec:proposed_enhancements}

\vspace*{-1mm}
\subsection{Proposed Consistency Checks}
\vspace*{-1mm}
As first contribution, we propose two novel consistency checks. Their aim is to achieve a reduced computational complexity and to increase the reconstruction quality.
\vspace*{-2mm}

\subsubsection{Fast Reverse Motion Check (\FaOMC{})}
\vspace*{-1mm}
The first proposed consistency check is related to \RME{}. Instead of calculating the reverse motion vector field, we test the more relevant motion vectors around the already found motion vector. In a first step, we therefore propose testing the same number of motions vectors as in \RME{} but placing them symmetrically around the motion vector pointing back to the original pixel. This is illustrated in Figure\,\ref{fig:consistency_checking}\,(b) and is further denoted as reverse motion check (\OMC{}). This algorithm can be assumed to be roughly as fast as \RME{} since the same number cost functions needs to be calculated.

In a second step, we propose testing only a tiny subset of these motion vectors leading to fast \OMC{} (\FaOMC{}). In case of an untrustworthy motion vector, the probability is high that many of the motion vectors in the reverse direction have a smaller cost than the currently chosen motion vector.
In our setup, we test only motions in the set $\{-7,-3,-1,0,1,3,7\}$ for both spatial dimensions instead of the testing all motion vectors $\{-9,-8,\dots, 9\}$  as in \OMC{} and \RME{}. The number of motions to be tested is significantly reduced from $361$ to $49$. 
An example is illustrated in Figure\,\ref{fig:consistency_checking}\,(c). If the green arrow has the lowest cost, the motion is accepted. If any of the red arrows has the lowest cost, the motion is rejected. %
\vspace*{-1mm}
\subsubsection{Nearest Neighbor Check (\FourNNC{})}
\vspace*{-1mm}
The second proposed consistency check does not require any additional template-matching at all and is therefore potentially faster. It is supposed to be used in combination with \FaOMC{} in order to speed up the calculations. This can be achieved since many motion vectors may already be sorted out with this simpler consistency check. It relies on testing the consistency of the determined vector field in a local neighborhood.
Denoting the found motion vector field as $(\alpha_{mn}, \beta_{mn})$, we perform a $3{\times}3$ median filtering for the two individual components resulting in the filtered motion vector field
\begin{align}
	(\tilde\alpha_{mn}, \tilde\beta_{mn}) = (\mathrm{median}_{3{\times}3}(\alpha_{mn}), \mathrm{median}_{3{\times}3}(\beta_{mn})).
\end{align}
Next, for each position $(m,n)$, the filtered motion vectors at the four nearest neighboring positions $(m-1,n), (m+1,n), (m,n-1)$, and  $(m,n+1)$ are compared. Only if the sum of the absolute differences of the motion vectors is at most one for each neighboring pair, the motion is accepted. If not, the motion is rejected because it is considered to be untrustworthy. Such accepted/rejected motions are highlighted with green/red color in Figure\,\ref{fig:consistency_checking}\,(d).
This modification of the consistency check is abbreviated as nearest neighbor-check (\FourNNC{}) later on. %

\vspace*{-1mm}
\subsection{Proposed Recursive FSR for Dynamic Masks (\RFSRmodified{})}
\vspace*{-1mm}
As second contribution to this paper, we propose a new implementation of a recursive FSR build upon the work from \cite{Jonscher2016a}. Other than \RFSR{} from \cite{Jonscher2016a}, our implementation handles the projected pixels differently and is capable of additionally handling dynamic masks. The novel algorithm is abbreviated as \RFSRmodified{}. 
During the model generation, we use the projected pixels in the same manner as \RFSR{}. As a last step, however, the  model found during the reconstruction is to be overwritten with the available measurements as commonly done in FSR \cite{Seiler2015}. In this step, \RFSR{} makes no difference between measured and projected pixels, whereas \RFSRmodified{} considers the projected pixels to be less reliable and therefore does not use them to overwrite the model.

\vspace*{-1mm}
\section{SIMULATIONS AND RESULTS}
\label{sec:simulation_and_results}
\vspace*{-1mm}
In this section, we evaluate the performance of the proposed consistency checks and \RFSRmodified{}. We compare them to \RFSR{} + \RME{} from \cite{Jonscher2016a}, investigate the impact of using a dynamic mask instead of a fixed mask, and show the runtimes. For any reconstructions with FSR, we chose the same parameters as in \cite{Jonscher2016a} except for the concealed weighting of the FSR being set to zero allowing us to perform a fully parallel processing of all blocks during the reconstruction. For all recursive reconstructions, we use three previous frames for the motion estimation and projection. For \RFSR{} + \RME{} from \cite{Jonscher2016a} we use raw simulation data kindly provided by the authors. This data is available for the fixed mask and one of the test sequence. For all other cases we use our own implementations as described in Section\,\ref{sec:proposed_enhancements}.

\begin{figure}[t]
	\centering
	{\footnotesize
\begingroup%
  \makeatletter%
  \providecommand\color[2][]{%
    \errmessage{(Inkscape) Color is used for the text in Inkscape, but the package 'color.sty' is not loaded}%
    \renewcommand\color[2][]{}%
  }%
  \providecommand\transparent[1]{%
    \errmessage{(Inkscape) Transparency is used (non-zero) for the text in Inkscape, but the package 'transparent.sty' is not loaded}%
    \renewcommand\transparent[1]{}%
  }%
  \providecommand\rotatebox[2]{#2}%
  \newcommand*\fsize{\dimexpr\f@size pt\relax}%
  \newcommand*\lineheight[1]{\fontsize{\fsize}{#1\fsize}\selectfont}%
  \ifx\svgwidth\undefined%
    \setlength{\unitlength}{244.69369507bp}%
    \ifx\svgscale\undefined%
      \relax%
    \else%
      \setlength{\unitlength}{\unitlength * \real{\svgscale}}%
    \fi%
  \else%
    \setlength{\unitlength}{\svgwidth}%
  \fi%
  \global\let\svgwidth\undefined%
  \global\let\svgscale\undefined%
  \makeatother%
  \begin{picture}(1,0.35445244)%
    \lineheight{1}%
    \setlength\tabcolsep{0pt}%
    \put(0.27342477,0.32306047){\color[rgb]{0,0,0}\makebox(0,0)[lt]{\lineheight{1.25}\smash{\begin{tabular}[t]{l}\textit{A}\end{tabular}}}}%
    \put(0.00797221,0.32583568){\color[rgb]{0,0,0}\makebox(0,0)[lt]{\lineheight{1.25}\smash{\begin{tabular}[t]{l}\textit{Spincalendar}\end{tabular}}}}%
    \put(0,0){\includegraphics[width=\unitlength,page=1]{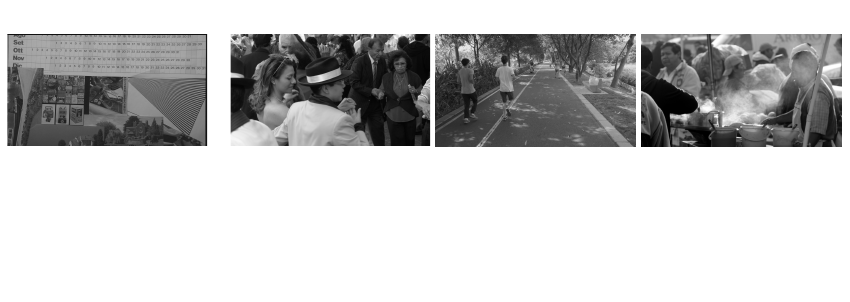}}%
    \put(0.00596971,0.14530673){\color[rgb]{0,0,0}\makebox(0,0)[lt]{\lineheight{1.25}\smash{\begin{tabular}[t]{l}\textit{Class C}\end{tabular}}}}%
    \put(0,0){\includegraphics[width=\unitlength,page=2]{datasets.pdf}}%
  \end{picture}%
\endgroup%

	}
	\vspace*{-5mm}
	\footnotesize
	\caption{Frame 20 of each video sequence used for the simulations.}
	\label{fig:testdata}
	\vspace*{-4mm}
\end{figure}   
For the test sets, we us several monochrome video sequences: The first 100 frames of the \textit{Spincalendar} sequence having a resolution of $1280{\times}720$ pixels are used since these were also used in \cite{Jonscher2016a}.
Further video data is taken from the  \textit{JVET} test sequences \cite{JVET-N1010}. For the  \textit{JVET -- ClassC} sequences, the resolution is  $832{\times}480$ pixels and we use the first 50 frames. Moreover, we chose three sequences from \textit{JVET -- A}. For those, we spatially down-scaled the frames by a factor of three resulting in $1280{\times}720$ pixels to achieve a similar resolution as for the other sequences. Once more, we use the first 100 frames. Figure\,\ref{fig:testdata} depicts a single frame of each used sequence.

To evaluate the quality of the reconstructed videos, we calculate the frame-wise PSNR and average it for all frames of the respective video. For the PSNR calculation, a border of 40 pixels is omitted since boundary effects are not considered to be of interest in our evaluation. The PSNR values are then further averaged across the video sequences to achieve a meaningful average value. The same evaluations were done for the mean structural similarity (SSIM)~\cite{Wang2004}.

\begin{figure*}[t]
	\centering
	{\footnotesize
		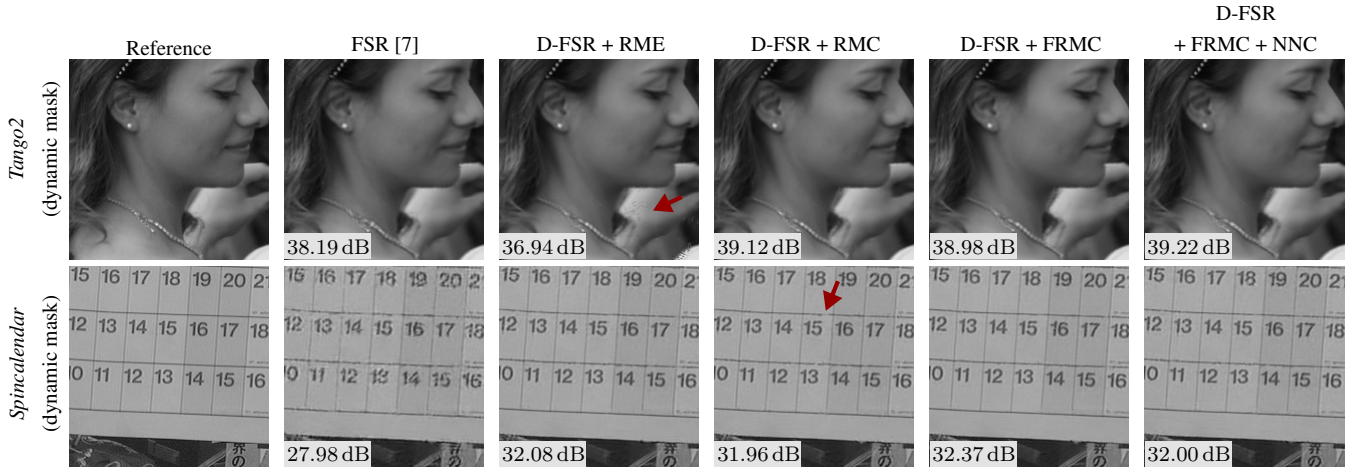
	}
	\vspace*{-5mm}
	\footnotesize
	\caption{Visual comparison of the different reconstruction methods for sections of frame 20 from the \textit{Tango2} sequence and frame 30 from the \textit{Spincalendar} sequence. The PSNR values given as insets are calculated for the visible sections. Since only dynamic masks are used, data for \RFSR{}+\RME{} \cite{Jonscher2016a} is not available. \textit{(Best to be viewed enlarged on a monitor.)}}
	\label{fig:compare_visual}
		\vspace*{-4mm}
\end{figure*}

\begin{table}[t]
	\footnotesize
	\caption{Results for the fixed mask. The average reconstruction quality in terms of PSNR in dB is provided using different test sets. For the averages, SSIM is provided, too.}
	\vspace*{-2mm}
	\scriptsize
	\label{tab:results_fixed}
	\centering
	\setlength{\tabcolsep}{3pt}
	\begin{tabularx}{0.99\linewidth}{ll||c|c|c|c|c|c}
		& \textit{\textbf{(fixed mask)}}                            &                   &       \RFSR{}        &                 &                 &                 & \RFSRmodified{} \\
		&                                                           &        FSR        &       + \RME{}       & \RFSRmodified{} & \RFSRmodified{} & \RFSRmodified{} &   + \FaOMC{}    \\
		&                                                           & \cite{Seiler2015} & \cite{Jonscher2016a} &    + \RME{}     &    + \OMC{}     &   + \FaOMC{}    &  + \FourNNC{}   \\ \hline
		& \rule{0pt}{1\normalbaselineskip}Spincalendar              &       30.38       &        31.66         &      32.67      &      32.72      & \textbf{33.00}  &      32.68      \\ \hline
		\rule{0pt}{1\normalbaselineskip}	\multirow{4}{*}{\rotatebox{90}{ \textit{Class C}}} & BasketballDrill                                           &       31.27       &          -           &      31.10      & \textbf{31.44}  &      31.38      & \textbf{31.44}  \\
		& BQMall                                                    &       27.49       &          -           &      27.73      &      27.81      & \textbf{27.82}  &      27.81      \\
		& PartyScene                                                &       23.22       &          -           &      23.35      &      23.37      &      23.37      & \textbf{23.39}  \\
		& RaceHorses                                                &       28.69       &          -           &      28.32      &      28.96      &      28.85      & \textbf{29.15}  \\ \hline
		\rule{0pt}{1\normalbaselineskip}	\multirow{3}{*}{\rotatebox{90}{ \textit{A}}}       & Tango2                                                    &       39.80       &          -           &      37.78      &      40.58      &      40.40      & \textbf{40.62}  \\
		& ParkRunning3                                              &       30.16       &          -           &      31.21      &      31.27      &      31.19      & \textbf{31.43}  \\
		& FoodMarket4                                               &       47.23       &          -           &      41.83      & \textbf{46.53}  &      45.72      &      46.35      \\ \hline
		& \rule{0pt}{1\normalbaselineskip}\textit{\textbf{Average (PSNR)}} &       32.28       &          -           &      31.75      &      32.83      &      32.72      & \textbf{32.86}\\\hline\hline
		& \rule{0pt}{1\normalbaselineskip}\textit{\textbf{Average (SSIM)}} &         0.9334 	& - &    0.9368 	 &    0.9399 	 &    0.9394 	 &    \textbf{0.9406} 
	\end{tabularx}
	\vspace*{-2mm}
\end{table}
\begin{table}[t]
	\vspace*{2mm}
	\footnotesize
	\caption{Results for the dynamic mask. The average reconstruction quality in terms of PSNR in dB is provided using different test sets. For the averages, SSIM is provided, too.}
	\vspace*{-2mm}
	\scriptsize
	\label{tab:results_vary}
	\centering
	\setlength{\tabcolsep}{3pt}
	\begin{tabularx}{0.9\linewidth}{ll||c|c|c|c|c}
		& \textit{\textbf{(dynamic mask)}}                          &                       &                 &                 &                 & \RFSRmodified{} \\
		&                                                           &                       & \RFSRmodified{} & \RFSRmodified{} & \RFSRmodified{} &   + \FaOMC{}    \\
		&                                                           & FSR \cite{Seiler2015} &    + \RME{}     &    +  \OMC{}    &   + \FaOMC{}    &  +  \FourNNC{}  \\ \hline
		& \rule{0pt}{1\normalbaselineskip}Spincalendar              &         30.43         &      33.20      &      33.26      & \textbf{33.56}  &      33.21      \\ \hline
		\rule{0pt}{1\normalbaselineskip}	\multirow{4}{*}{\rotatebox{90}{ \textit{Class C}}} & BasketballDrill                                           &         31.30         &      33.62      &      34.28      & \textbf{34.37}  &      34.31      \\
		& BQMall                                                    &         27.53         &      30.01      &      30.14      & \textbf{30.39}  &      30.27      \\
		& PartyScene                                                &         23.23         &      24.64      &      24.70      & \textbf{25.05}  &      24.92      \\
		& RaceHorses                                                &         28.71         &      28.34      &      29.02      &      28.92      & \textbf{29.22}  \\ \hline
		\rule{0pt}{1\normalbaselineskip}	\multirow{3}{*}{\rotatebox{90}{ \textit{A}}}       & Tango2                                                    &         39.84         &      37.59      &      40.69      &      40.51      & \textbf{40.73}  \\
		& ParkRunning3                                              &         30.19         &      31.36      &      31.42      &      31.34      & \textbf{31.58}  \\
		& FoodMarket4                                               &         47.28         &      41.56      & \textbf{46.58}  &      45.78      &      46.38      \\ \hline
		& \rule{0pt}{1\normalbaselineskip}\textit{\textbf{Average (PSNR)}} &         32.31         &      32.54      &      33.76      &      33.74      & \textbf{33.83}\\\hline\hline
		& \rule{0pt}{1\normalbaselineskip}\textit{\textbf{Average (SSIM)}} &         0.9338 	 &    0.9477 	 &    0.9510 	 &    0.9520 	 &    \textbf{0.9524}
	\end{tabularx}
	\vspace*{-2mm}
\end{table}

Tables\,\ref{tab:results_fixed} and \ref{tab:results_vary} show the results of the reconstruction quality in terms of PSNR using a fixed and a dynamic mask, respectively.
Besides the results using the single frame FSR \cite{Seiler2015}, the various consistency checks are shown in combination with \RFSRmodified{}. Additionally, the SSIM was evaluated in the same manner. Its results are in accordance with the PSNR values and the averages are provided in the last row of Tables\,\ref{tab:results_fixed} and \ref{tab:results_vary} for completeness.
Comparing the average results from both tables, we find that using a dynamic mask outperforms using a fixed mask by roughly \SI[retain-explicit-plus]{+1}{dB} which is consistent with the findings in \cite{Jonscher2018}.

In Table\,\ref{tab:results_fixed}, we can observe that \RFSRmodified{} + \RME{}, outperforms the original version from \cite{Jonscher2016a} by \SI[retain-explicit-plus]{+1.01}{dB} for the \textit{Spincalendar} sequence. Beyond this, we investigated the influence of the proposed consistency checks on the reconstruction quality in terms of PSNR.
Using \OMC{}, the reconstruction quality averaged over all used sequences is increased by  \SI[retain-explicit-plus]{+1.08}{dB} for the fixed mask and \SI[retain-explicit-plus]{+1.22}{dB} for the dynamic mask.
Using the fast variant of \OMC{}, namely \FaOMC{}, results in a slight decrease of the average PSNR. Interestingly, this trend is not uniform across the different sequences. For example, the \textit{FoodMarket4} scene shows a relevant loss whereas the reconstruction for other sequences improves. The average loss is reasonable, as not all motion vector are tested in the opposite direction.
Lastly, the \FourNNC{} is added to the \FaOMC{}. This combination shows the highest average reconstruction quality in both Tables\,\ref{tab:results_fixed} and \ref{tab:results_vary}. For the dynamic mask, a gain of \SI[retain-explicit-plus]{+1.52}{dB} is observed compared to the single frame FSR \cite{Seiler2015} and a gain of \SI[retain-explicit-plus]{+1.29}{dB} is observed compared to \RFSRmodified{} + \RME{}. This means, that \FaOMC{} + \FourNNC{} overcomes the loss arising from switching to \FaOMC{} and even improves the quality in average.

\begin{figure}[t]
	\centering
	{\footnotesize
		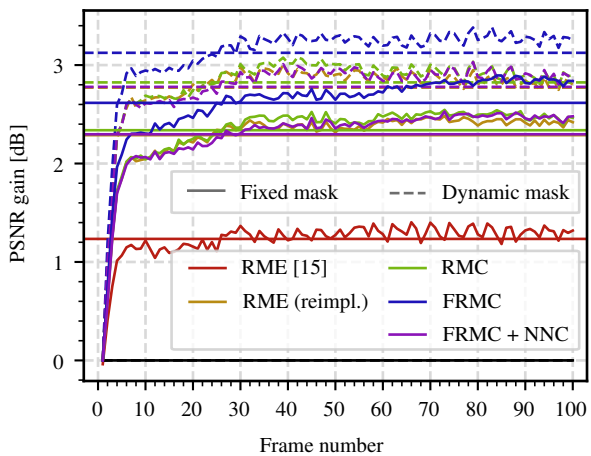
	}
	\vspace*{-4mm}
	\footnotesize
	\caption{Frame-wise  gain of the reconstruction quality in terms of PSNR relative to the single frame FSR \cite{Seiler2015} using the \textit{Spincalendar} sequence.}
	\label{fig:compare_Jonscher_vs_frames}
	\vspace*{-2mm}
\end{figure}

For a more in-depth view of the simulated data, Figure\,\ref{fig:compare_Jonscher_vs_frames} shows the frame-wise PSNR gain relative to the single frame FSR. The  \textit{Spincalendar} sequence is used. It can be seen that roughly 10 frames are needed for the recursive algorithm to converge to a good quality and that the gain is then mostly constant for all the remaining frames.

In order to be able to judge the visual quality, Figure\,\ref{fig:compare_visual} shows two sections of the reconstructed frames using the dynamic mask. For the example from the \textit{Tango2} sequence, it can clearly be seen that the rather low average PSNR of \RFSRmodified{} + \RME{} arises from strong artifacts indicated by the red arrow. These arise from faulty motion vectors that can occur in nearly constant regions in combination with large motion vectors and should be sorted out. For the other consistency checks, these motion vectors are sorted out as desired.
For the \textit{Spincalendar} sequence, the differences among the proposed algorithms are more subtle. The cases where \OMC{} performs slightly worse than \FaOMC{} + \FourNNC{} are highlighted with red arrows.

In addition to the reconstruction quality, we evaluate the computation times for the different algorithms in case of the dynamic masks. We provide the runtimes for the motion estimation, the consistency checks between the current frame and its last three frames, and the reconstruction. For the motion estimation a fast variant using the GPU was developed, too. For the motion estimation and consistency checks, we restrict the executions to a single core on an Intel i9-10980XE CPU with 3.00GHz. Table\,\ref{tab:runtime} summarizes the results. Timings for \RFSR{} + \RME{} from \cite{Jonscher2016a} are not available but its algorithmic complexity is identical to that of  \RFSRmodified{} + \RME{}.

The code for the pixel-wise motion estimation is identical for all three cases and therefore the results are all close. The same is true for the FSR. The slight differences can be used as an estimate of the accuracy of the measurements and are considered to be acceptable. Taking a look at the runtimes of the consistency checks, we can see that \FourNNC{} + \FaOMC{} is more than 13-fold faster than the \RME{} and \OMC{}.  Remarkably, combining \FaOMC{} and \FourNNC{}, is more than 8-fold faster than using only \FaOMC{} since many motion vectors can be sorted out using solely the very fast \FourNNC{}. In such cases, the slower \FaOMC{} is skipped. It is worth taking these times into relation with the total runtimes of the reconstruction, where an overall reduction of  \SI{-48}{\%} is achieved.

\begin{table}[t]
	\footnotesize
	\vspace*{2mm}
	\caption{Runtimes of the different steps in seconds. The algorithmic complexity of \RFSR{} + RME{} from \cite{Jonscher2016a} is identical to that of \RFSRmodified{} + \RME{}. ME: Motion estimation, CC: consistency check.}
	\scriptsize\vspace*{-2mm}
	\label{tab:runtime}
	\centering
	\setlength{\tabcolsep}{3pt}
	\begin{tabularx}{0.7\linewidth}{l||c|c|c||c}
		&  ME   &  CC   & \, FSR\, & Total \\ \hline
		\rule{0pt}{1\normalbaselineskip}\RFSRmodified{} + \RME{}  & 21.56 & 33.80 &   6.02   & 61.37 \\
		\RFSRmodified{} + \OMC{}                                            & 23.11 & 40.19 &   6.11   & 69.41 \\
		\RFSRmodified{} + \FaOMC{}                                          & 21.49 & 21.02 &   6.05   & 48.56 \\
		\RFSRmodified{} + \FourNNC{} + \FaOMC{}                             & 23.14 & 2.56  &   6.07   & \textbf{31.77}
	\end{tabularx}
	\vspace*{-3mm}
\end{table}

\vspace*{-1mm}
\section{CONCLUSION}
\vspace*{-1mm}
\label{sec:conclusion}
Using recursive reconstruction algorithms, a pixel-wise motion estimation and projection between the current frame and its preceding frames is performed to enhance the reconstruction quality of the current frame. Since some motion vectors may be untrustworthy, it is required to perform a consistency check which sorts out such motion vectors. For this task, \RFSR{} from \cite{Jonscher2016a} relies on a computationally expensive reverse motion estimation (\RME{}).
In order to reduce the cost, we propose a new consistency check which is a combination of \FaOMC{} and \FourNNC{}.  Altogether, more relevant reverse motion vectors are tested in \FaOMC{} and most evaluations are skipped by comparing the locally neighboring motion vectors using \FourNNC{}. The proposed \RFSRmodified{} uses  the projected pixels differently and can handle dynamic masks as well.

With our proposed recursive reconstruction method and consistency checks, \RFSRmodified{} + \FaOMC{} + \FourNNC{}, we achieve a \SI[retain-explicit-plus]{+1.01}{dB} higher reconstruction quality in terms of PSNR compared to \RFSR{} + \RME{} from \cite{Jonscher2016a} in case of the fixed mask and the \textit{Spincalendar} sequence. Testing a larger dataset of different video sequences, we find that the \RFSRmodified{} + \FaOMC{} + \FourNNC{}  performs better than \RFSRmodified{} + \RME{} by \SI[retain-explicit-plus]{1.29}{dB} in average for the dynamic mask.
The average PSNR gain with respect to the single frame FSR \cite{Seiler2015} is \SI[retain-explicit-plus]{+1.52}{dB}. 
At the same time, the proposed consistency check, is 13-fold faster than \RME{} which reduces the total runtime  by \SI{-48}{\%}.

\bibliographystyle{IEEEbib}
\bibliography{literatur_jabref}

\end{document}